# RENORMALIZATION

# OF THE

# CLASSICAL VELOCITY

# IN THE

# LATTICE HEAVY QUARK EFFECTIVE THEORY

*Jeffrey E. Mandula*
*U.S. Department of Energy*
*Division of High Energy Physics*
*Washington, DC  20585, USA*

*and*

*Michael C. Ogilvie*
*Department of Physics*
*Washington University*
*St. Louis, MO  63130, USA*

# ABSTRACT

In the lattice formulation of the Heavy Quark Effective Theory (LHQET), the "classical velocity" $v$ becomes renormalized. The origin of this renormalization is the reduction of Lorentz (or O(4)) invariance to (hyper)cubic invariance. The renormalization is finite, depends on the form of the decretization of the reduced heavy quark Dirac equation, and can persist down to zero lattice spacing. For the Forward Time - Centered Space discretization, the renormalization is computed both perturbatively, to one loop, and non-perturbatively, using an ensemble of lattices with $\beta = 6.1$ provided by the Fermilab ACP-MAPS Collaboration. The estimates of the leading multiplicative shift agree reasonably well, and indicate that for small classical velocities, $\vec{v}$ is reduced by about 20-25%.

# I    INTRODUCTION

In the Isgur-Wise heavy quark limit[1,2], there are many relations between the decay constants and form factors of particles containing a heavy quark. Among the most striking of these conclusions from the heavy quark spin-flavor symmetry is the fact that in the $M_Q \to \infty$ limit, a single form factor, the Isgur-Wise universal function $\xi$, describes all semileptonic decays of one meson containing a heavy quark into another, such as the process $B \to D^* l \nu$. From the first, it has been emphasized that although the heavy quark spin-flavor symmetry suffices to infer the existence of this function, its calculation requires non-perturbative techniques, such as lattice gauge theory[3]. Several such calculations have been carried out. A calculation by the present authors used a lattice implementation of the heavy quark effective theory[4], and two other lattice calculations treat the heavy quarks as Wilson fermions with a small hopping constant, but do not directly implement the heavy quark limit[5,6].

On the lattice[4], as in the continuum[1,2,7], the Isgur-Wise limit entails the introduction of a "classical velocity", $v$, normalized to 1, which appears in the decomposition of the momentum of a heavy particle and the reduced Dirac equation of the heavy quark field:

$$P = Mv + p$$
$$-iv \cdot D\, h^{(v)}(x) = 0 \tag{1}$$

In the continuum, the velocity that appears in these two contexts is the same. However, on the lattice this is not the case. In this article we explain the origin of this new renormalization and describe two calculations of its magnitude. One calculation is a one-loop perturbative calculation. This follows the analysis of Aglietti[8] on a variant lattice Dirac operator. The second calculation is non-perturbative. It is based on a computation of the shift in the energy of a meson containing a heavy quark, as measured by the fall-off of its Euclidean space propagator, for a given shift in its residual momentum.

The incorporation of the above $M \to \infty$ limit into a Lagrangian is accomplished by factoring out a phase which is singular in the $M \to \infty$ limit and defining a reduced field[7]

$$\frac{1 + \gamma \cdot v}{2} h^{(v)}(x) = e^{-iMv \cdot x} \frac{1 + \gamma \cdot v}{2} \psi(x) \qquad (2)$$

In the limit, the Lagrangian for $h^{(v)}(x)$ becomes

$$\mathcal{L}^{(v)} = \bar{h}^{(v)}(x) \, iv \cdot D \, h^{(v)}(x) = \bar{h}^{(v)}(x) \, iv \cdot [\partial - igA(x)] \, h^{(v)}(x) \qquad (3)$$

The propagator of the reduced field $h^{(v)}(x)$ is

$$S^{(v)}(p) = \frac{1}{v \cdot p - \Sigma^{(v)}(p)} \qquad (4)$$

where in perturbation theory $\Sigma^{(v)}(p)$ is the sum of all 1-particle irreducible self mass diagrams, as illustrated by Figure 1.

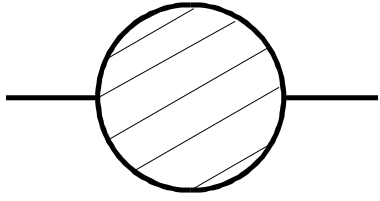

Figure 1 — The proper self mass shift of the reduced heavy quark propagator

## II      RENORMALIZATION OF THE CLASSICAL VELOCITY

If $\Sigma^{(v)}(p)$ is Taylor expanded in powers of the residual momentum

$$\Sigma^{(v)}(p) = m + \left.\frac{\partial \Sigma^{(v)}}{\partial p_\mu}\right|_{p=0} p_\mu + \cdots \tag{5}$$

the first term gives a mass shift. This is without physical significance and can be eliminated by a redefinition of the reduced heavy quark field by a non-singular phase which is independent of the heavy mass[9].

$$h^{(v)}(x) \to e^{imv\cdot x} h^{(v)}(x) \tag{6}$$

This corresponds to a finitely different breakup of the total 4-momentum

$$P = Mv + p \rightarrow Mv' + (p + mv) \tag{7}$$

In the heavy quark limit, $v$ and $v'$ become equal. The renormalized classical velocity is inferred from the small residual momentum behavior of the full heavy quark propagator

$$S^{(v)}(p) = \frac{Z^{(v)}}{-m + v^{(ren)} \cdot p + \cdots} \tag{8}$$

This identifies the renormalized classical velocity as:

$$v_\mu^{(ren)} = Z^{(v)}(v_\mu - X_\mu)$$

$$X_\mu \equiv \left.\frac{\partial \Sigma^{(v)}}{\partial p_\mu}\right|_{p=0} \tag{9}$$

The wave function renormalization constant $Z^{(v)}$ is determined by the requirement that the classical velocity has unit (Minkowski space) normalization both before and after the shift

$$v_\mu^{(ren)\,2} = v_\mu^2 = Z^{(v)\,2}(v_\mu - X_\mu)^2 = -1 \tag{10}$$

In the continuum, $X_\mu$ must be proportional to the only available 4-vector, $v_\mu$. Therefore, it modifies the inverse of the heavy quark propagator to first order in $p$ only by an overall multiplicative factor, which amounts to a wave function renormalization of $h^{(v)}(x)$, but gives no velocity shift.

On the lattice, because of the reduced Lorentz or rotational symmetry, $X_\mu$ need not be proportional to $v_\mu$. For example, it can have a term proportional to $v_\mu^3$, which has the same O(4) transformation properties as $v_\mu$, but is linearly independent of it. When the coefficient $X_\mu$ is not simply proportional to $v_\mu$, the classical velocity is shifted.

An equivalent procedure for calculating the renormalized classical velocity, but which is better suited to non-perturbative simulation, is to consider the rate of fall-off of the reduced propagator in Euclidean time for fixed spacial momentum.

$$S(t,\vec{p}) \sim Z^{(v)} e^{-E^{(v)}(\vec{p}) t} \tag{11}$$

Allowing for a possible residual mass, the propagator decay rate is

$$\begin{aligned} E^{(v)}(\vec{p}) &= \lim_{M \to \infty} \sqrt{(M+m)^2 + ((M+m)\vec{v}^{(phys)} + \vec{p})^2} - M\sqrt{1 + \vec{v}^{(phys)2}} \\ &= m v_0^{(phys)} + \frac{\vec{v}^{(phys)} \cdot \vec{p}}{v_0^{(phys)}} \\ & \left( v_0^{(phys)} = \sqrt{1 + \vec{v}^{(phys)2}} \right) \end{aligned} \tag{12}$$

To help keep track of the context, we will use the phrase the physical classical velocity, $v_\mu^{(phys)}$, to refer to the renormalized classical velocity, when it is defined non-perturbatively through the rate of decay of a propagator in Euclidean time.

The logarithmic derivative of the asymptotic propagator

$$\frac{\partial S^{(v)}(t,\vec{p})/\partial p_i |_{\vec{p}=0}}{S^{(v)}(t,\vec{p}=0)} \sim -\frac{\partial E^{(v)}(\vec{p})}{\partial p_i}\bigg|_{\vec{p}=0} t$$

$$= -\frac{v_i^{(phys)}}{v_0^{(phys)}} t \equiv -\tilde{v}_i^{(phys)} t \quad (13)$$

directly extracts the physical classical velocity.

$$v_\mu^{(phys)} = \frac{(1,\tilde{v}^{(phys)})}{\sqrt{1-\tilde{v}^{(phys)^2}}} \quad (14)$$

We now proceed to two calculations of the renormalized classical velocity  The first uses one-loop perturbation theory, and the second uses a lattice simulation. In both, we use the forward time - centered space discretization of the heavy quark Dirac equation[4]

$$v_0 [ U_4(x,x+\hat{t}) \tilde{S}^{(v)}(x+\hat{t},y) - \tilde{S}^{(v)}(x,y) ]$$
$$+ \sum_{\mu=1}^{3} \frac{-iv_\mu}{2} [U_\mu(x,x+\hat{\mu}) \tilde{S}^{(v)}(x+\hat{\mu},y) - U_\mu(x,x-\hat{\mu}) \tilde{S}^{(v)}(x-\hat{\mu},y) ] = \delta(x,y) \quad (15)$$

### III    ONE-LOOP RENORMALIZATION OF THE CLASSICAL VELOCITY

The one loop the proper self mass is given by the diagrams shown in Figure 2. The point interaction only gives rise to a residual mass, and so can be ignored. With an infrared regulator $\lambda$, the second diagram gives

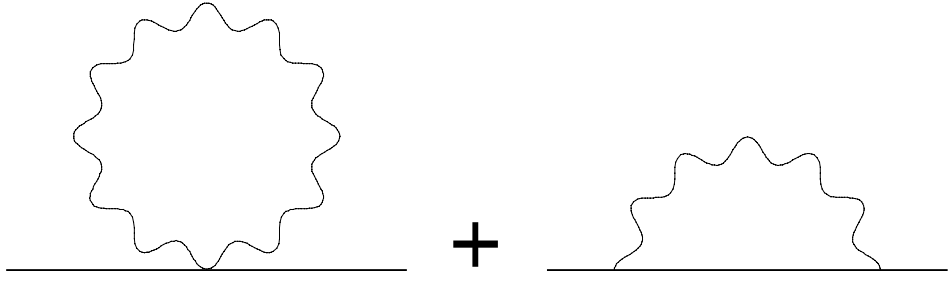

Figure 2 — The 1-loop contribution to the heavy quark proper self mass

$$\Sigma^{(2)} = g^2 C_2 \int \frac{d^4 k}{(2\pi)^4} V_\mu(p - 2k) \Delta_{(\lambda)}(k) S^{(v)}(p - k) V_\mu(p - 2k) \quad (16)$$

The integration domain is periodic, $[-\pi/a, \pi/a]$, and the bare propagators and vertices are

$$S^{(v)^{-1}}(p) = \frac{v_0}{a}\left(e^{ip_4 a} - 1\right) + \sum_i \frac{v_i}{a} \sin p_i a$$

$$\Delta_{(\lambda)}^{-1}(k) = \frac{4}{a^2} \sum_\mu \sin^2 \frac{k_\mu a}{2} + \lambda^2 \quad (17)$$

$$V_\mu^a(q = 2p + k) = g t^a \left(i v_0 e^{i q a/2}, \; v_i \cos \frac{q_i a}{2}\right)$$

The factor of the quadratic Casimir invariant, $C_2$, which is $(N^2 - 1)/2N$ for SU($N$) color with quarks in the fundamental representation, arises from the color contraction in the loop integral.

The Taylor series coefficient of $p_\mu^1$ is a non-linear function of $v_\mu$ and is furthermore asymmetric between space and time (because of the asymmetry of the lattice action, which is a centered difference in space but a forward difference in time). Thus there is a classical velocity shift as well as a multiplicative renormalization constant $Z^{(v)}$.

As has been noted, the symmetrical first difference results in fermion doubling. However, unlike in the usual fermion situation, the contributions of the secondary modes automatically vanish in the zero spacing limit. The reason for this is that each is accompanied by an undoubled gluon at the edge of the Brillouin zone, whose energy is on the order of the inverse lattice spacing[8].

The evaluation of the loop integral over the gluon 4-momentum requires care in the choice of contour in the complex Euclidean energy plane. The path of the contour with respect to the poles in the propagators is determined by the coordinate space boundary conditions. The contour must be chosen so that the heavy quark propagator always vanishes for negative Euclidean time. Thus the propagator as a function of momentum and Euclidean time is given by

$$\int \frac{dE}{2\pi} \frac{e^{iEt}}{\frac{v_0}{a}[e^{iEa} - 1] + \frac{1}{a}\sum_i v_i \sin p_i a}$$

$$= \frac{1}{2\pi i} \oint \frac{dz}{z} \frac{z^{(t/a)}}{v_0 [z - 1] + \sum_i v_i \sin p_i a}$$

$$(z = e^{iEa})$$

(18)

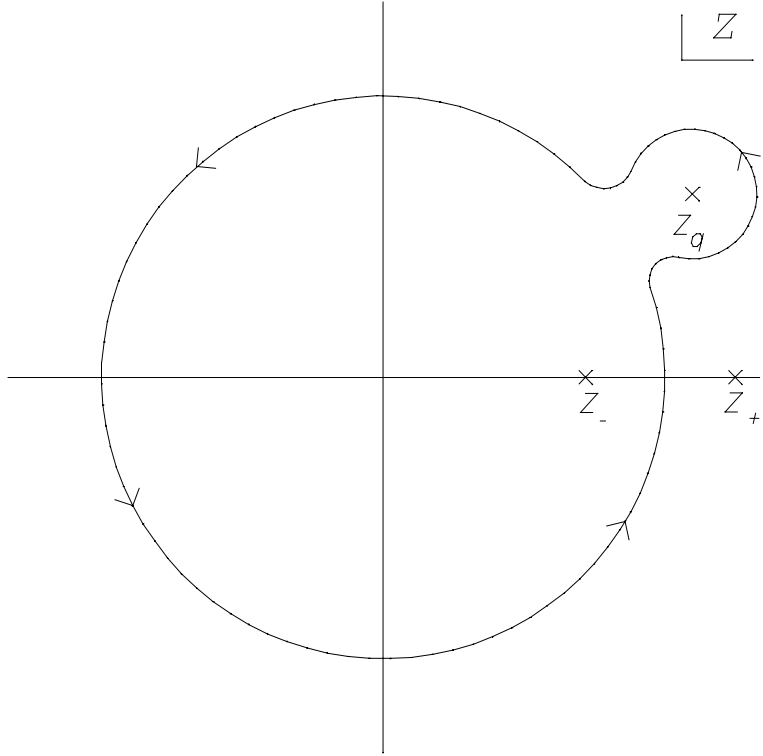

Figure 3 — Contour in the $z = e^{iEa}$ plane for the one loop proper self mass

and the necessity for this to vanish for negative $t$ no matter the values of $v$ and $p$ implies that the contour in the $z$ plane always encloses the pole.

Although 3-momentum integrals will eventually have to be done numerically, it is necessary to do the energy integration analytically. With the notation

$$z = e^{iak_4}$$

$$A = 4 \sum_{i=1}^{3} \sin^2 \frac{k_i a}{2} + (\lambda a)^2$$

$$B = \sum_{i=1}^{3} v_i \sin(p_i + k_i) a \qquad (19)$$

$$C = \sum_{i=1}^{3} v_i^2 \cos^2\left(p_i + \frac{k_i}{2}\right) a$$

the Euclidean energy integral is

$$\frac{a^2}{2\pi i} \oint \frac{dz}{z} \frac{-v_0^2 e^{2ip_4 a} z + C}{[-(z - 2 + 1/z) + A][v_0(e^{ip_4 a} z - 1) + B]} \qquad (20)$$

where the integration is around a closed contour in the $z$ plane. Note that the quantities $A$ and $C$ are both positive, but $B$ can take either sign. The singularities of the integrand are located at

$$z_\pm = 1 + \frac{A}{2} \pm \frac{1}{2}\sqrt{4A + A^2}$$

$$z_q = e^{-ip_4 a}\left(1 - \frac{1}{v_0} B\right) \qquad (21)$$

One of the pair of poles coming from the gluon propagator is within the unit circle in the $z$ plane, and the other is outside. Depending on the sign of $B$, the pole coming from the heavy quark propagator can be either inside or outside. The contour must be chosen to pass between the two gluon propagator poles, which is the ordinary proceedure. However, the requirement that the heavy quark propagator must vanish for all negative Euclidean times means that the the contour must always enclose the quark propagator pole, whatever the sign of $B$.[10] The appropriate contour is shown in Figure 3, for the non-standard case, negative $B$.

The integrand decays sufficiently rapidly at infinity so that the energy integral is given by the residue of the single pole outside the contour, at $z_+$. The resulting expression for the one loop proper self mass is

$$\Sigma^{(2)} = g^2 C_2 \frac{a^2}{(2\pi)^3} \int d^3k \frac{-v_0^2 e^{2ip_4 a} z_+ + C}{\sqrt{4A + A^2} \left[v_0 (e^{ip_4 a} z_+ - 1) + B\right]} \tag{22}$$

To facilitate comparison with the evaluation of the renormalization of the classical velocity by means of simulation, we express the perturbative result in terms of $\tilde{v}_i = v_i / v_0$, which can be seen from the reduced heavy quark Dirac equation (Eq. (15)) to be the actual transverse hopping expansion parameter of the lattice HQET. Its lowest order perturbative renormalization is given by

$$\delta \tilde{v}_i \equiv \tilde{v}_i^{(ren)} - \tilde{v}_i = -\frac{1}{v_0}(X_i - \tilde{v}_i X_0) \tag{23}$$

Although this infrared finite expression can be numerically evaluated for any value of the classical velocity, explicitly evaluating the leading terms in a series on the components of $\tilde{v}_i$ brings out the structure of the result:

$$\delta \tilde{v}_i = c_1 \tilde{v}_i + c_3 \tilde{v}_i^3 + c_{12} \tilde{v}_i \sum_{j \neq i} \tilde{v}_j^2 + \cdots \tag{24}$$

The expansion coefficients are given by

$$c_1 = \frac{-g^2 C_2}{(2\pi)^3} \int d^3\theta \, \frac{z_+(\cos\theta_i + z_+ - 2)}{\sqrt{4A + A^2}\,(z_+ - 1)^2}$$

$$c_3 = \frac{-g^2 C_2}{(2\pi)^3} \int d^3\theta \, \frac{(\cos\theta_i + 1)(z_+ - 3\cos\theta_i + 2)(z_+^2 + 2z_+(\cos\theta_i - 2) + 1)}{2\sqrt{4A + A^2}\,(z_+ - 1)^4}$$

$$c_{12} = \frac{-g^2 C_2}{(2\pi)^3} \int d^3\theta \, \frac{(\cos\theta_j + 1)[2z_+(z_+ - 3\cos\theta_i + 2)\cos\theta_j - (z_+^2 - 8z_+ + 1)\cos\theta_i + z_+(z_+^2 - 4z_+ - 3)]}{2\sqrt{4A + A^2}\,(z_+ - 1)^4}$$

$$(j \neq i) \tag{25}$$

where the integration is over the periodic box $\theta_i \in [-\pi, \pi]$.

The numerical values of these coefficients, for $g^2 = 6/\beta = 6/6.1$ are

$$c_1 = -.23345568$$
$$c_3 = -.04143250 \qquad (26)$$
$$c_{12} = -.03881061$$

The difference between the values of the two cubic terms is a reflection of the reduced spatial symmetry of the lattice. The perturbative evaluation of the expansion coefficients of the corrections to the classical velocity are shown in Table II, for both naive and Lepage-Mackensie[11] tadpole improved mean field shifted values of the coupling.

## IV     SIMULATION OF THE CLASSICAL VELOCITY RENORMALIZATION

To evaluate the renormalization of the classical velocity non-perturbatively, we must be somewhat indirect. Without global gauge fixing, we cannot simulate the quark propagator. However, if we form a heavy-light meson, its classical velocity will be that same as that of its heavy quark component, and its propagator is easily computable in simulations. The calculation is greatly facilitated by noting that because the solution of the heavy quark lattice Dirac equation (15) is obtained by forward recursion, after n time steps the propagator is an $(n-1)^{st}$ order polynomial in the bare classical velocity components $\tilde{v}_i = v_i/v_0$.

$$S^{(v)}(t,\vec{p}) = \sum_{m_1,m_2,m_3} \tilde{v}_1^{m_1} \tilde{v}_2^{m_2} \tilde{v}_3^{m_3} S(t,\vec{p},\vec{m}) \tag{27}$$

The computation of the coefficients in this polynomial is highly efficient. The significance of the indices $m_i$ is that they are the maximum lattice displacement of the heavy quark propagator in the $i^{th}$ direction. The only essential difference between the asymptotic behavior of the heavy quark propagator and that of a heavy-light meson is the possibility, on the lattice, that the overall asymptotic normalization of the meson field composite operator depends on the residual momentum.

$$S^{(v)}(t,\vec{p}) \sim C^{(v)}(\vec{p}) e^{-E^{(v)}(\vec{p}) t} \tag{28}$$

The slope of the logarithmic derivative of the asymptotic propagator versus $t$,

$$\begin{aligned}
\frac{\partial S^{(v)}(t,\vec{p})/\partial p_i |_{\vec{p}=0}}{S^{(v)}(t,\vec{p}=0)} &\sim \frac{\partial C^{(v)}(\vec{p})/\partial p_i |_{\vec{p}=0}}{C^{(v)}(\vec{p}=0)} - \left.\frac{\partial E^{(v)}(\vec{p})}{\partial p_i}\right|_{\vec{p}=0} t \\
&= \frac{\partial C^{(v)}(\vec{p})/\partial p_i |_{\vec{p}=0}}{C^{(v)}(\vec{p}=0)} - \tilde{v}_i^{(phys)} t
\end{aligned} \tag{29}$$

is the negative of the physical classical velocity. Since actual lattices are finite, we use the lattice approximation to the momentum derivative

$$\left.\frac{\partial S^{(v)}(t,\vec{p})}{\partial p_i}\right|_{\vec{p}=0} \Rightarrow \left.\frac{\Delta S^{(v)}(t,\vec{p})}{\Delta p_i}\right|_{\vec{p}=0}$$

$$\equiv \frac{1}{2p_{\min}}\left[S^{(v)}(t, p_i = p_{\min}) - S^{(v)}(t, p_i = -p_{\min})\right] \quad (30)$$

$$\left(p_{\min} = \frac{2\pi}{Na}\right)$$

The shift in the classical velocity to a given order $n$ in the components of the bare classical velocity $\tilde{v}_i = v_i/v_0$, only depends on the heavy quark propagator at values of the spatial coordinates up to $n$ lattice sites away from the initial heavy quark location.

$$\left.\frac{\Delta S^{(v)}(t,\vec{p})}{\Delta p_i}\right|_{\vec{p}=0,\,O(v_i^n)} = \frac{Na}{4\pi} \sum_{|x_i| \leq n} \left(2i \sin\frac{2\pi x_i}{N}\right) \times S^{(v)}(t,\vec{x}) \quad (31)$$

As with the perturbative evaluation, we display the results of the simulation to 3$^{\text{rd}}$ order in the bare classical velocity. The coefficients of powers of $\tilde{v}_i^{m_i}$ in the logarithmic derivative are

$$\begin{aligned}
\frac{\Delta S^{(v)}(t,\vec{p})/\Delta p_i \vert_{\vec{p}=0}}{S^{(v)}(t,\vec{p}=0)} &= \sum_i \tilde{v}_i \, \frac{\Delta S(t,\vec{p},m_i=1)/\Delta p_i \vert_{\vec{p}=0}}{S(t,\vec{p}=0,\vec{m}=0)} \\
&+ \sum_i \tilde{v}_i^3 \left[ \frac{\Delta S(t,\vec{p},m_i=3)/\Delta p_i \vert_{\vec{p}=0}}{S(t,\vec{p}=0,\vec{m}=0)} \right. \\
&\quad \left. - \frac{\Delta S(t,\vec{p},m_i=1)/\Delta p_i \vert_{\vec{p}=0} \; S(t,\vec{p}=0,m_i=2)}{S(t,\vec{p}=0,\vec{m}=0)^2} \right] \\
&+ \sum_{i,j\ne i} \tilde{v}_i \tilde{v}_j^2 \left[ \frac{\Delta S(t,\vec{p},m_i=1,m_j=2)/\Delta p_i \vert_{\vec{p}=0}}{S(t,\vec{p}=0,\vec{m}=0)} \right. \\
&\quad \left. - \frac{\Delta S(t,\vec{p},m_i=1)/\Delta p_i \vert_{\vec{p}=0} \; S(t,\vec{p}=0,m_j=2)}{S(t,\vec{p}=0,\vec{m}=0)^2} \right] \\
&+ \cdots
\end{aligned} \quad (32)$$

Averaging over directions of course improves the signal to noise ratio.

We have simulated these propagator expansion coefficients on an ensemble of lattices and Wilson light quark propagators made available to us by the Fermilab ACP-MAPS Collaboration[12]. These consisted of 98 lattices of size $24^3 \times 48$ with lattice coupling $\beta = 6.1$ along with Wilson quark propagators with hopping constant $\kappa = .154$. We computed the coefficients of the leading terms in the $\tilde{v}_i$ expansion of the heavy quark propagator, and formed heavy-light meson propagators (or, more precisely, their $\tilde{v}_i$ expansion coefficients) in the standard manner. We did not use smearing or otherwise improve the saturation of the resulting objects by the lowest mass state contribution, but it is our intention to use systematic variational methods to improve the isolation of the ground state in a future, more refined analysis [12, 13].

The simulation did take advantage of the fact that, since the direction of time is conventional, purely forward, or purely backward propagation of the heavy quarks in Euclidean time gave effectively independent ensembles of lattices and propagators.

Figure 4 shows the results of the simulation of the heavy-light meson propagator. The solid circles are the values of the leading, zeroth order term in the expansion of the propagator in powers of $\tilde{v}_i$, which is the static propagator. Also shown are a single exponential fit to the static propagator with a fall-off given by the average efffective mass over the range between 2 to 6 units of Euclidean time, and the fractional saturation of the propagator by this single exponential. The other propagator components shown are those coefficients of the linear and cubic terms in the the $\tilde{v}_i$ expansion of the propagator that enter into the simulation of $\tilde{v}^{(phys)}$ through third order in the bare classical velocity components, that are shown in Eq. (32). Table I shows the fitted slope of each coefficient.

The degree to which a single exponential saturates the propagator is an indication of how well the ground state is isolated. The fact that there good saturation for at most a very small range of Euclidean times shows that by the time one has reached sufficiently large Euclidean separations to isolate the ground state, the statistical errors in the propagator have become overwhelming. This is an independent way of seeing that there is not an extended region of Euclidean time over which the effective mass of heavy-light propagator is constant[14]. A plot of the effective mass of the $\vec{m} = 0$ propagator in each Euclidean time interval is shown in Figure 5. The absence of a stable plateau must be remediated either by better statistics, or probably more effectively by improving the heavy-light meson wave function.

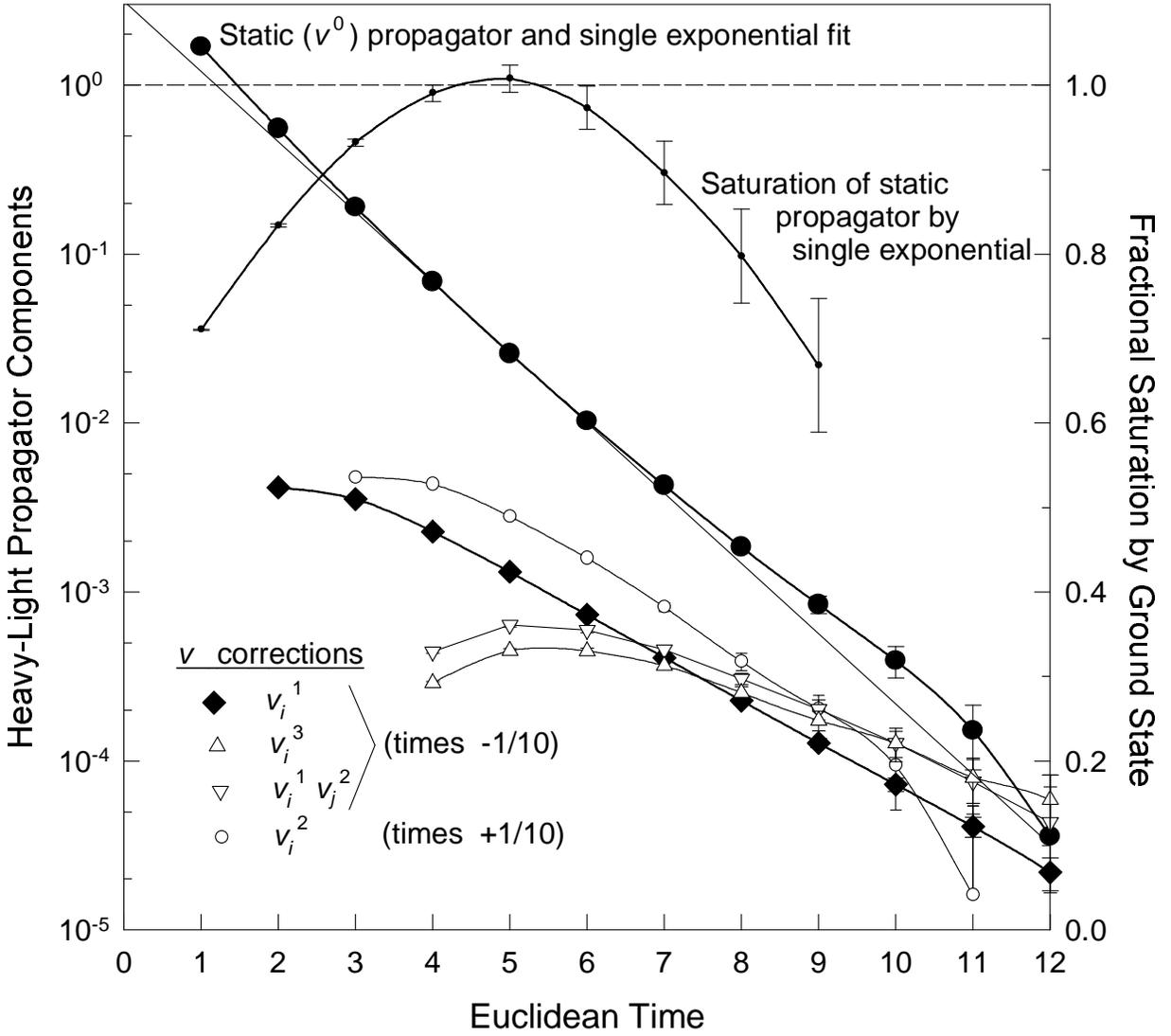

Figure 4 — The coefficients of powers of the bare classical velocity in the heavy-light meson propagator through third order

The three combinations of ratios of propagator components that enter into the third order evaluation of the physical classical velocity shown in Eq. (32) are shown in Figure 6. The errors are the propagated statistical errors associated with the individual terms in the propagator expansion. As shown by Eqs. (29) and (32), the asymptotic slopes of these combinations of propagator contributions are the coefficients of the indicated terms in the expansion of $-\tilde{v}_i^{(phys)}$ in powers of $\tilde{v}_i^{m_i}$. In order to estimate them, we fit each coefficient function to a straight line, starting at $t = 3$ for the linear coefficient and at $t = 4$ for the cubic coefficients. The fitted slopes are shown in Table I.

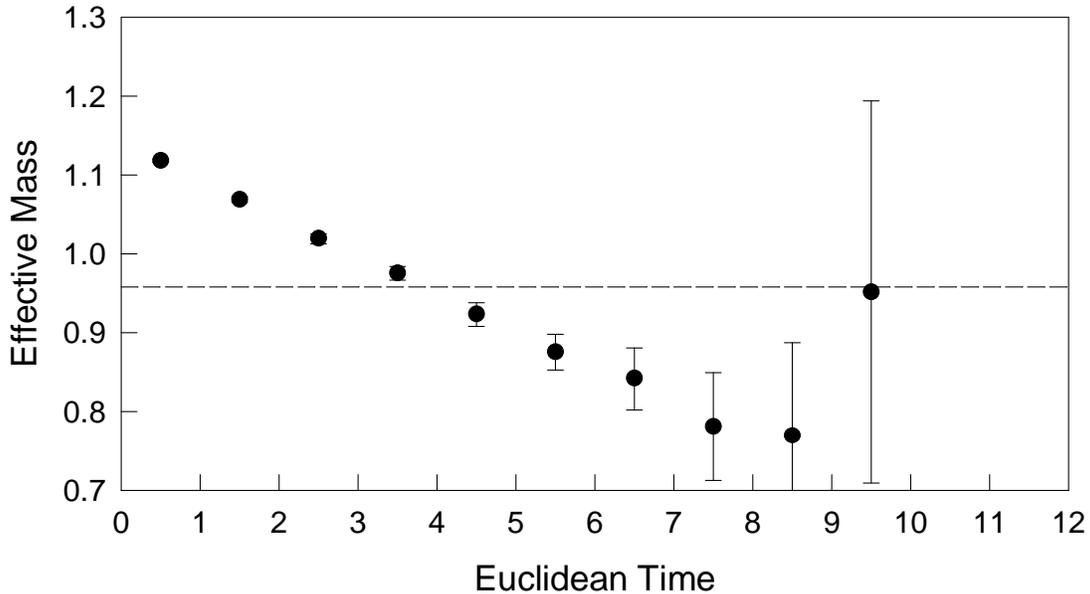

Figure 5 — The effective mass of the zero classical velocity, zero residual momentum heavy-light meson propagator between each time slice

## V  TADPOLE IMPROVEMENT

In the analysis so far, we have not taken advantage of the tadpole improvement proceedure of Lepage and Mackensie[11]. This consists in simply replacing each link variable

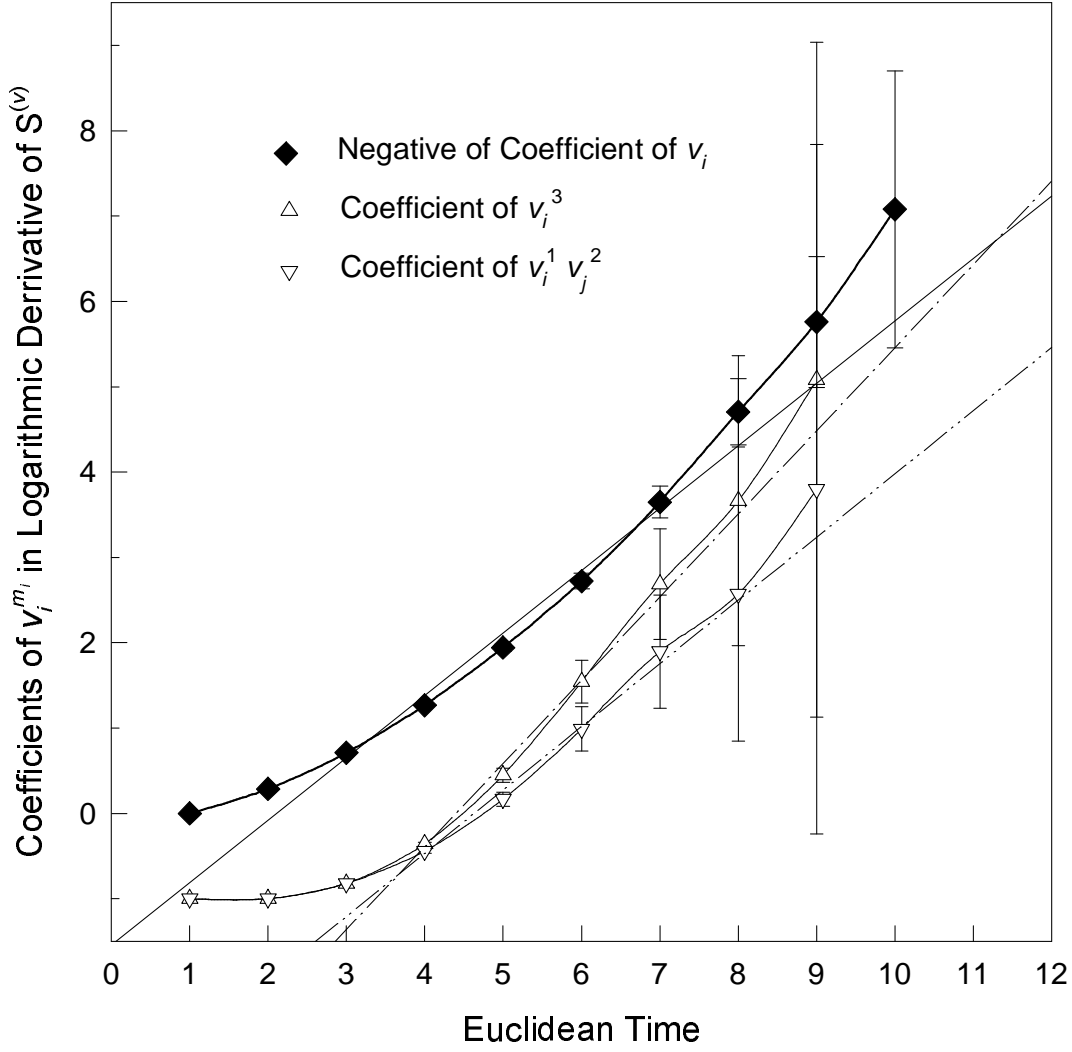

Figure 6 — The three combinations of propagators that are the coefficients of the linear and cubic terms (in $v$) in the logarithmic derrivative of $S^{(v)}$, as in Eq. (32)

$U_\mu(n)$ by that link divided by its average value $U_\mu(n)/u_0$, where the average value of the link is given by the gauge-invariant expression

Table I - Fitted slopes from simulation of heavy-light meson propagator

| Coefficient of | Fitted Slope | Fitting Range |
| --- | --- | --- |
| $v_i$ | $-.713 \pm .092$ | 3 - 9 |
| $v_i^3$ | $.982 \pm .087$ | 4 - 8 |
| $v_i v_j^2 \ (i \neq j)$ | $.764 \pm .078$ | 4 - 8 |

$$u_0 = \left[\frac{1}{3} \langle Tr \Box \rangle \right]^{1/4} \tag{33}$$

where $\Box$ denotes the product of the link variables around an elementary $1 \times 1$ plaquette. In the present context there are two aspects to this proceedure. One is the identification of the appropriate perturbative coupling associated with a simulation at a given value of $\beta$, which would simply be $6/g^2$ without tadpole improvement. For the Wilson action used to generate the Fermilab lattices,

$$I = -\frac{6}{g^2} \sum_\Box \frac{1}{3} Tr \Box \Rightarrow -\frac{6}{g^2 u_0^4} \sum_\Box \frac{1}{3} Tr \Box \tag{34}$$

$\beta$ instead is identified with $6/g^2 u_0^4$. Since $u_0$ is necessarily less than one, this has the effect of increasing the value of the effective perturbative coupling corresponding to a lattice simulation at a given value of $\beta$.

The other aspect is the improvement of the heavy quark propagator. If each link in the discretized reduced Dirac equation (15) is replaced by itself divided by $u_0$, then modulo an overall normalization, which of course does not affect the determination of the physical classical velocity, the effect of all the factors of $u_0$ can be absorbed into a redefinition of the lattice bare classical velocity and a mass renormalization. Specifically, if we absorb a factor of $u_0^{-n_t} = \exp(-n_t \ln u_0)$ into the heavy quark propagator, we recover the original discretized equation (Eq. (15)) for the modified propagator function, but with the replacement

$$\tilde{v}_i = \frac{v_i}{v_0} \Rightarrow \frac{\tilde{v}_i}{u_0} \tag{35}$$

The factor of $u_0^{-n_t}$ is evidently just a shift in the mass by $-\ln u_0$. As with any other residual mass, it is without physical effect. The replacement in the classical velocity means that it is actually the quantity $\tilde{v}_i / u_0$ which is the transverse hopping constant. The factors of $u_0$ that come from replacing $\tilde{v}_i$ by $\tilde{v}_i / u_0$ can be absorbed into the coefficients of powers of the bare classical velocity $\tilde{v}_i$ in the expansion of the physical classical velocity $\tilde{v}_i^{(phys)}$. The coefficient of a term with $N$ powers of $\tilde{v}_i$, whether determined by simulation or perturbatively, is multiplied by $u_0^{-N}$. In terms of the expansion coefficients of $\delta \tilde{v}$ of Eq. (24), if we denote the improved coefficients by $\hat{c}$, the relation between the improved and unimproved sets of coefficients is

$$\hat{c}_1 = \frac{c_1 + 1 - u_0}{u_0}$$

$$\hat{c}_3 = \frac{c_3}{u_0^3}$$

$$\hat{c}_{12} = \frac{c_{12}}{u_0^3} \tag{36}$$

$$\hat{c}_{ijk} = \frac{c_{ijk}}{u_0^{i+j+k}} \qquad (i+j+k > 1)$$

Table II — Comparison of Simulated and Perturbative Expansion Coefficients

| Coefficient | Term | Naive Coupling | | Tadpole Improved | |
|---|---|---|---|---|---|
| | | Simulation | One-Loop | Simulation | One-Loop |
| $c_1$ | $v_i$ | -.287 ± .092 | -.2334 | -.186 ± .105 | -.3113 |
| $c_3$ | $v_i^3$ | -.982 ± .087 | -.0414 | -1.46 ± .130 | -.1048 |
| $c_{12}$ | $v_i v_j^2$ ($i \neq j$) | -.764 ± .078 | -.0388 | -1.14 ± .116 | -.0982 |

## V  DISCUSSION OF THE RESULTS

It is striking to compare the perturbative evaluation of these expansion coefficients with the results of the simulation performed at $\beta = 6.1$. This is shown in Table II, both with and without tadpole improvement. The coefficients of the linear term agree rather well between the simulated and perturbative evaluations. There is a statistically insignificant preference for the use of the naive rather than the tadpole improved coupling in this context. However, the perturbative and simulated cubic coefficients are in complete disagreement. It is difficult to surmise which estimate, if either, is correct. While it is possible that the lattice coupling is simply too large to neglect higher order terms, is should be pointed out that the simulations of the cubic (and higher) coefficients are very delicate calculations.

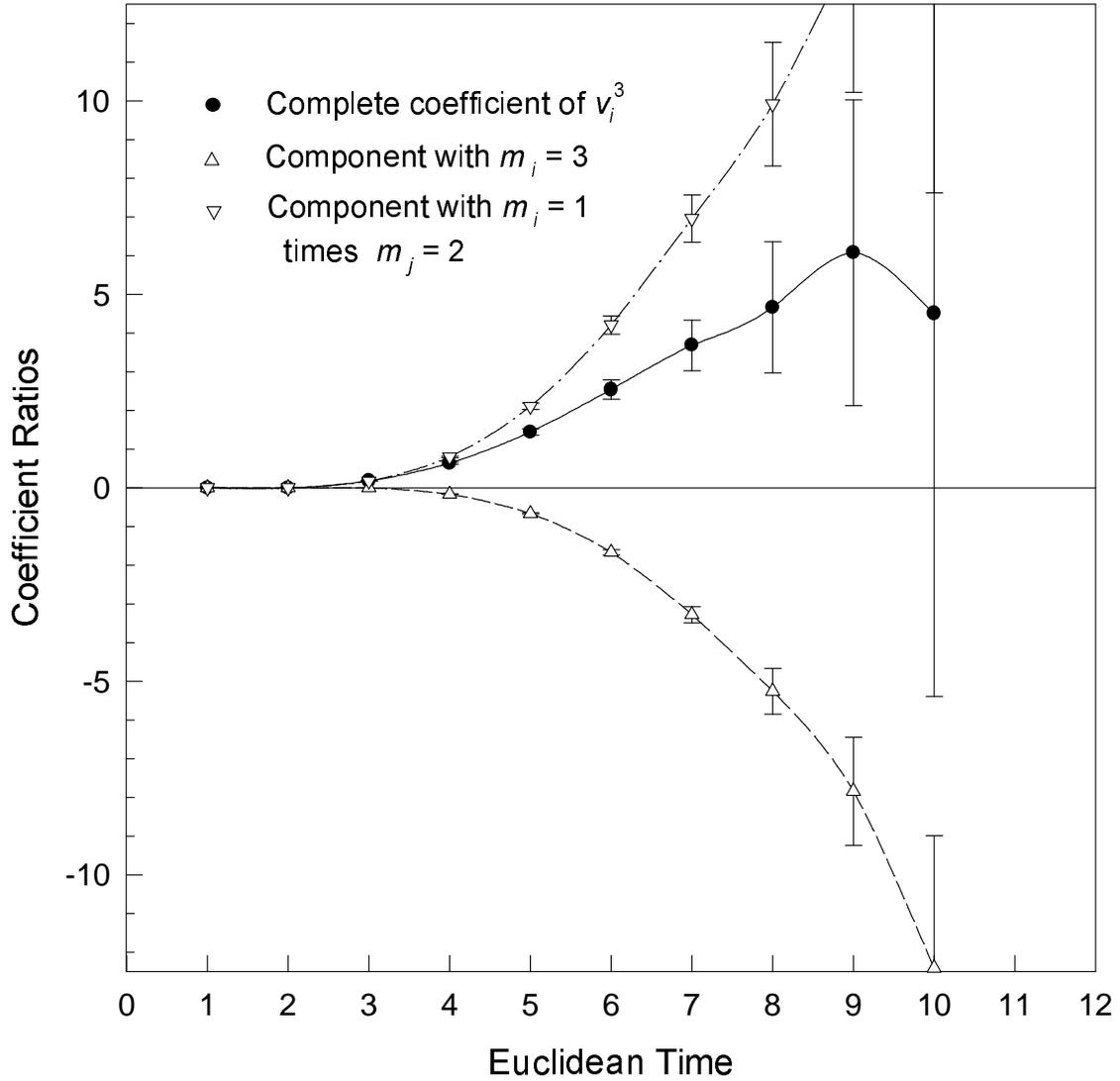

Figure 7 — Composition of the $v_i^3$ term in the heavy-light meson propagator

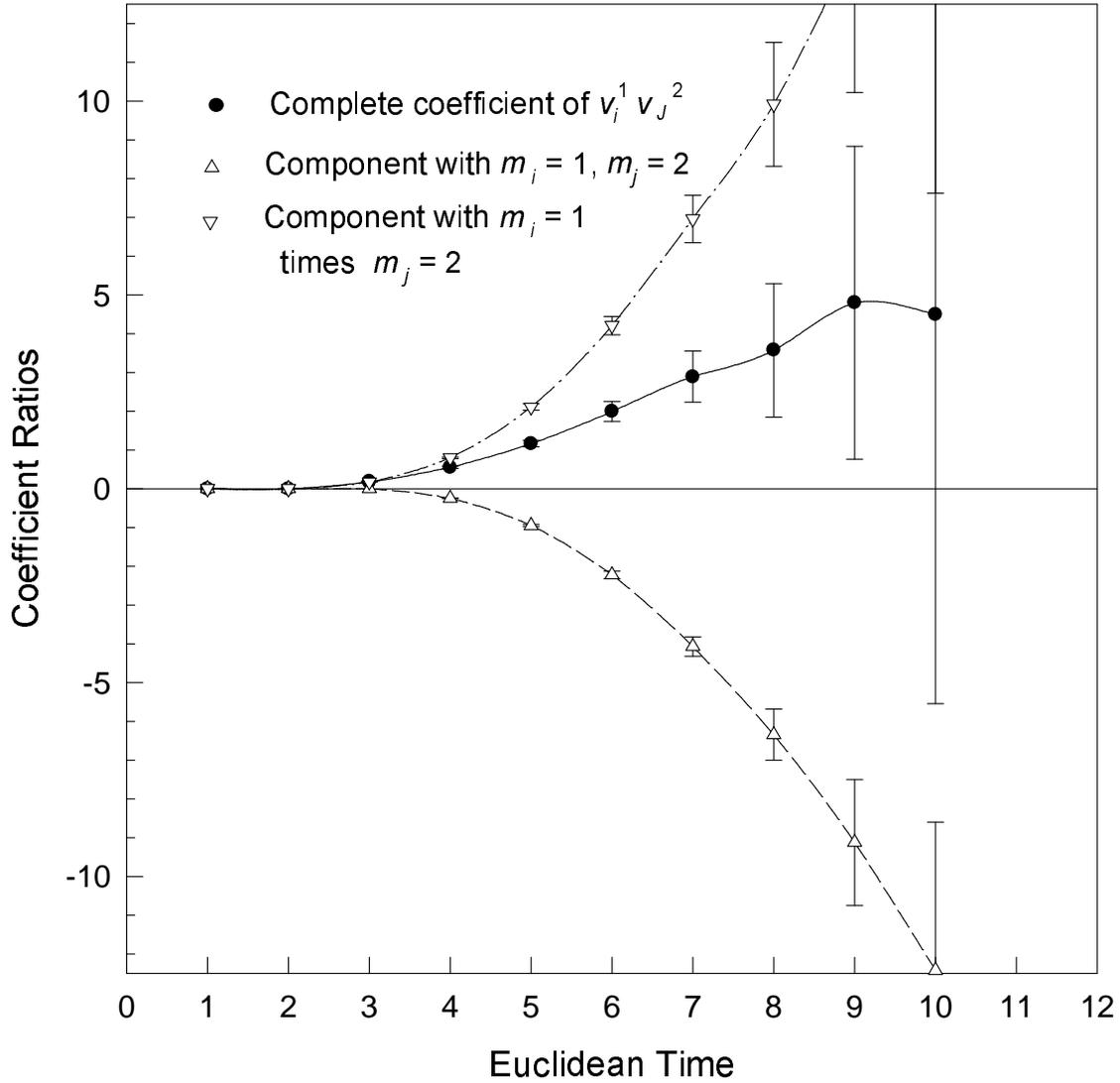

Figure 8 — Composition of the $v_i^1 \, v_j^2$ term in the heavy-light meson propagator

Each of the cubic coefficients in the expansion of the logarithmic derivative of the heavy-light meson propagator is the difference of two terms, each of which is substantially larger than their difference. Furthermore, each of the two terms has a $t^3$ asymptotic behavior, while their difference should only grow linearly. Thus as one tries to eliminate non-asymptotic artifacts by going to large $t$, the statistical quality of the calculation rapidly breaks down. Fortunately, the dominant, linear coefficient is free of this particular difficulty. The large cancellations are roughly of the same magnitude in both cubic coefficients. The degree of cancelation is shown for the $v_i^3$ term in Figure 7 and for the $v_i v_j^2$ term in Figure 8. Note that the $t^3$ asymtotic behavior of the individual terms is clearly visible from the smallest values of the Euclidean time.

Let us sum up the results of these analyses of the renormalization of the classical velocity in lattice version of the heavy quark effective theory. The origin of this new renormalization is the reduction of the Lorentz invariance of the continuum to hypercubic symmetry on the lattice. Both in perturbation theory and via simulations the shift in the classical velocity is quite substantial. The most reliable results concern the first-order, multiplicative shift. The two analyses are in fairly good agreement on this point. Since the lattice coupling corresponds to a continuum coupling constant near 1, it is probably most reasonable to use the result coming from the simulation, which is intrinsically non-perturbative. This first-order (in $\tilde{v}_i^{(bare)}$) multiplicative renormalization is

$$\frac{\tilde{v}_i^{(phys)} - \tilde{v}_i^{(bare)}}{\tilde{v}_i^{(bare)}} \cong \begin{cases} -.186 \pm .105 & (\text{Tadpole improved}) \\ -.287 \pm .092 & (\text{Unimproved}) \end{cases} \tag{37}$$

In contrast to the first-order situation, the simulated and the perturbative third-order results differ from each other by an order of magnitude. Neither seems to us terribly reliable, though it is hard to invent a reason for the perturbative result to be an order of magnitude off. By contrast, the substantial intrinsic errors we have described above in the simulation of the cubic coefficients make it quite conceivable that the asymptotic slope has not yet been observed in the simulations reported here. Since the perturbative result for these coefficients is tiny, one may hope that for moderate values of the classical velocity, the first-order correction suffices for practical use.

**ACKNOWLEDGEMENTS**

The authors would like to thank the Fermilab ACP-MAPS group for generously making available the lattice configurations and propagators that were used in the non-perturbative evaluation of the renormalization of the classical velocity.

contour correctly expresses the physics of the HQET. The discretization of the heavy quark reduced Dirac equation with a symmetric time derivative is not a valid lattice implementation of the heavy quark limit.